\documentclass{aa}
\usepackage{times}
\usepackage{natbib}
\usepackage{graphicx,longtable,cancel}
\usepackage{caption}
\usepackage{subfig}
\usepackage{booktabs}
\usepackage{tablefootnote}
\usepackage{geometry}
\usepackage{float,lscape}
\usepackage{xspace}
\usepackage{verbatim} 
\usepackage{txfonts}
\usepackage{supertabular}

\bibpunct{(}{)}{;}{a}{}{,}

\def\nh{$N_{\rm H}$\xspace}

\def\ltsima{$\; \buildrel < \over \sim \;$}
\def\lsim{\lower.5ex\hbox{\ltsima}}
\def\gtsima{$\; \buildrel > \over \sim \;$}
\def\gsim{\lower.5ex\hbox{\gtsima}}

\def\ngc{NGC\,6540\xspace}
\def\src{J1806--27\xspace}
\def\xmm{{\em XMM--Newton}\xspace}

\def\cha{{\em Chandra}\xspace}

\begin{document}

\title
{EXTraS discovery of a peculiar    flaring X-ray source in~the Galactic globular cluster NGC 6540}

\author{Sandro Mereghetti\inst{1}, Andrea De Luca\inst{1,4}, David Salvetti\inst{1},  Andrea Belfiore\inst{1}, Martino Marelli\inst{1,3}, Adamantia Paizis\inst{1}, Michela Rigoselli\inst{1,2}, Ruben Salvaterra\inst{1},    Lara Sidoli\inst{1}, Andrea Tiengo\inst{3,4,1}}

\institute {INAF, Istituto di Astrofisica Spaziale e Fisica
Cosmica Milano, via E.\ Bassini 15, I-20133 Milano, Italy
\and
Dipartimento di Fisica G. Occhialini, Universit\`a degli Studi di Milano Bicocca, Piazza della Scienza 3, I-20126 Milano, Italy
\and
IUSS, Piazza della Vittoria 15, I-27100 Pavia, Italy
\and
INFN, Sezione di Pavia, via A. Bassi 6, I-27100 Pavia, Italy
  }

\offprints{sandro@iasf-milano.inaf.it}

\date{Received / Accepted}

\authorrunning{S. Mereghetti et al.}

\titlerunning{ }

\abstract{ We report the discovery of a   flaring X-ray source in the globular cluster \ngc , obtained during the EXTraS project devoted to a systematic search for variability in archival data of the \xmm satellite. The source had a quiescent X-ray  luminosity of the order of $\sim$10$^{32}$ erg s$^{-1}$   in the  0.5--10 keV range  (for a distance of \ngc of 4 kpc) and showed a flare lasting about 300 s.  During the flare, the X-ray luminosity increased by more than a factor 40,  with a total emitted energy of $\sim$10$^{36}$ erg.  These properties, as well as {\em Hubble} Space Telescope photometry of the possible optical counterparts, suggest the identification with a chromospherically active binary. However, the flare luminosity is significantly higher than  what  commonly observed in stellar flares of such a short duration, leaving open the possibility of other interpretations.
\keywords{globular clusters: individual (\ngc ) -- stars: flares -- X-rays: binaries }}

\maketitle
 
\section{Introduction}

Globular clusters host a large number of X-ray sources.  
The luminous ones  ($L_{\rm X}>10^{36}$ erg s$^{-1}$) were soon recognized as low mass X-ray binaries (LMXB) containing neutron stars, due to the detection of type I X-ray bursts  in many of them and  to  their overall properties similar to those of LMXBs in the field \citep{1984ApJ...282L..13G}.  
The nature of the much larger population of sources with lower X-ray luminosity has been more difficult to understand. It is now clear that low luminosity sources ($<10^{35}$ erg s$^{-1}$) in globular clusters comprise   a mix of different classes, including  compact objects powered by accretion and/or rotation (transient LMXBs in quiescence, cataclysmic variables, millisecond pulsars), as well as non-degenerate stars with enhanced  X-ray emission related to chromospheric    activity \citep[see, e.g.,][]{2010AIPC.1314..135H}.
  
Here we report the discovery of a transient X-ray source with peculiar variability properties in the globular cluster \ngc . This result was obtained  in the course of   EXTraS\footnote{http://www.extras-fp7.eu/}, an EU/FP7 project devoted to a systematic variability study of the X-ray sources in the   \xmm\ public archive \citep{2016ASSP...42..291D}.

  \ngc is located in the Galactic bulge and distances between 3 and 5.3 kpc have been derived by various authors (see \citealt{2015MNRAS.450.3270R}, and references therein). In the following we give all the distance-dependent quantities normalized to an assumed value $d_4 = d / \rm{4 kpc}$.

\begin{figure*}
	\centering
\includegraphics[width=14cm]{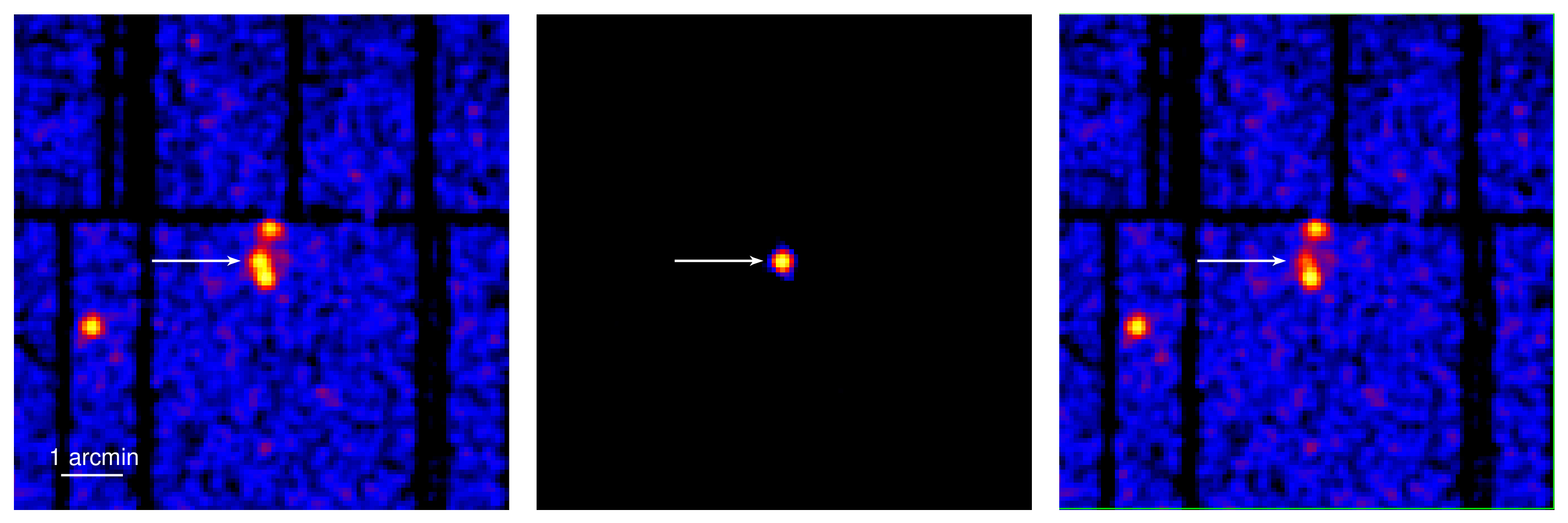}
	\caption{X-ray images of a 8$'$$\times$8$'$  field toward \ngc (North to the top, East to the left). The images are obtained by summing the data of the three EPIC cameras in the 0.2-12 keV energy range. Left panel: whole observation, with an exposure time of about 9 ks. Center panel: a 300 s time interval centered on the flare. 
Right panel: whole observation with the above time interval excluded. The source indicated by the arrow  is clearly variable:  the majority of its counts are detected during the short time interval containing the flare, but the source is visible also when the time interval containing the  flare is excluded.
\label{fig-ima}}
 \end{figure*}

\section{Data analysis and results}

\subsection{\xmm}

\ngc was observed with \xmm\ for about 9 ks  on 2005 September 21, starting at   03:09 UT. 
Our results are based on  data obtained with the three CCD cameras of the EPIC instrument. During this observation, all of them used the medium thickness optical filter and were operated in full imaging mode, resulting in a time resolution of  73 ms for the pn camera \citep{2001A&A...365L..18S} and  2.6 s for the two MOS cameras \citep{2001A&A...365L..27T}.

\begin{figure}[ht!]
	\centering
\includegraphics[width=7.7cm]{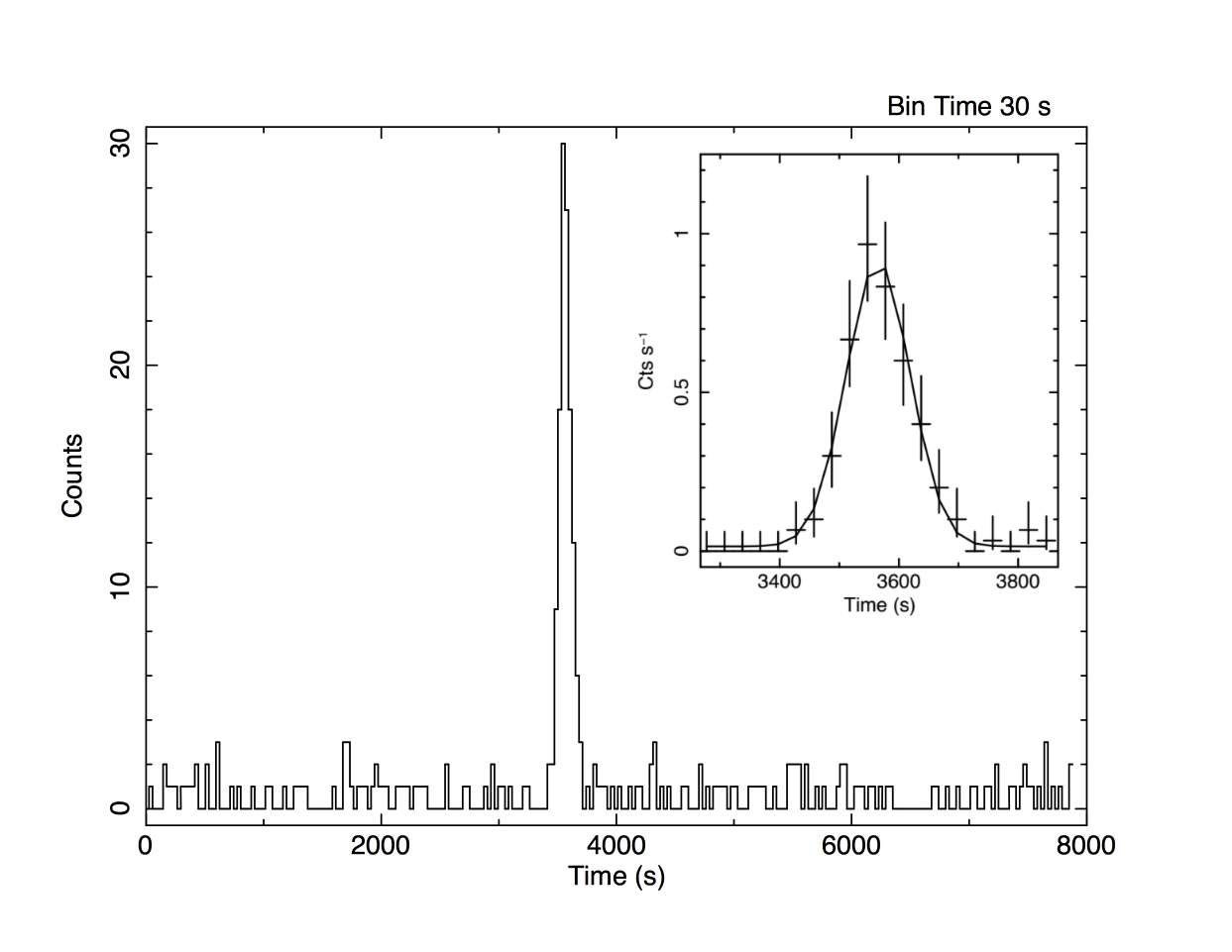}
	\caption{X-ray light curve (PN+MOS) of the transient source \src with a bin time of 30
s. Start time is set at the beginning of the pn observation (MJD=53634.14811
-- all times are in Barycentric Dynamical Time). The inset shows a zoom on
the flare. Error bars are computed according to \citet{1986ApJ...303..336G}. The
flare is well described by a Gaussian centered at
MJD=$53634.18937\pm0.00014$ and with $\sigma=54\pm11$ s.
\label{fig-lc}}
 \end{figure}

The left panel of  Fig.~\ref{fig-ima} shows a 8$'$$\times$8$'$ region of the 0.2-12 keV image obtained by summing the data of the three EPIC cameras. This image, integrated over the whole observation,  shows the presence of  four sources.  The variable source, 
at coordinates    R.A. = 18$^h$  6$^m$ 8.9$^s$, Dec. = $-27^{\circ}$ 45$'$  53$''$,  is indicated by the arrow. Its light curve is shown in Fig.~\ref{fig-lc}, where a flare  lasting about 300 s is clearly visible.

This source is also present in the 3XMM Catalog of \xmm serendipitous  sources \citep{2016A&A...590A...1R}, with the name 3XMM~J180608.9--274553. In the following we will refer to it with the abbreviated name \src . 
 
The middle panel of Fig.~\ref{fig-ima} shows  the same sky region accumulated during the short time interval (300 s) corresponding to the flare.
src is  detected with  high significance in this image,  which is too short to reveal the other sources, but it is also 
visible  when the flare time interval is excluded, as shown in the right panel of  Fig.~\ref{fig-ima}.

\setlength{\tabcolsep}{1em}
\begin{table*}[htbp!]
\centering \caption{Results of the spectral fits of \src }
\label{tab-spectra}
\begin{tabular}{lcccc}
\toprule 
Model	      & \nh	                                 & $\Gamma$ / kT		&      Unabsorbed  flux (0.5-10 keV)            & $\chi_{\nu}^2$/d.o.f.	\\
		      & $10^{21}$~cm$^{-2}$  	&       keV	                 &   erg cm$^{-2}$ s$^{-1}$    & 	          \\
\hline \\
\multicolumn{5}{c}{Flare emission} \\
\hline \\
Power law   &    6$\pm$2   	                & 1.7$\pm$0.2         	  & (3.6$^{+0.4}_{-0.3})\times10^{-12}$	  &  0.78/9 	\\[2pt]
Blackbody$^{a}$  &     $<$1.3                       &     1.0$\pm$0.1        	   &  	 ($2.3\pm0.2)\times10^{-12}$            	& 0.52/9	\\[2pt]
Bremsstrahlung  &   5$^{+2}_{-1}$    &     9.7$^{+11.4}_{-3.8}$   &  ($3.3\pm0.3)\times10^{-12}$     	& 0.69/9	\\[2pt]
Mekal         &    6$\pm$2   	         &   6.6$^{+11.5}_{-2.4}$     	&  	 ($3.4\pm0.3)\times10^{-12}$                & 0.76/9 	\\[2pt]
Diskbb         &    4$\pm$1   	         &   1.9$^{+0.4}_{-0.3}$     	&  	 ($2.8\pm0.3)\times10^{-12}$                & 0.54/9 	\\[2pt]
\hline \\
\multicolumn{5}{c}{Quiescent emission} \\
\hline \\
Power law   &    6 (fixed)  	                & 2.5$\pm$0.5         	& ($10\pm2)\times10^{-14}$   	& 	0.2/6\\[2pt]
Blackbody$^{a}$  &     0.3 (fixed)             &    0.7$\pm$0.1          	&  	 ($4\pm1)\times10^{-14}$        	&  0.37/6	\\[2pt]
Bremsstrahlung  &   5 (fixed)             &   2.2$^{+1.0}_{-0.6}$         &  	         ($7\pm1)\times10^{-14}$            	& 0.29/6	\\[2pt]
Mekal         &     6 (fixed)   	         &     2.2$^{+0.6}_{-0.4}$         	&    ($8\pm1)\times10^{-14}$                    	& 	0.20/6\\[2pt]
Diskbb         &    4 (fixed)  	         &   0.9$\pm$0.2     	&  	 ($7\pm1)\times10^{-14}$                & 0.31/6	 \\[2pt]

\bottomrule\\
\end{tabular}

\raggedright
\scriptsize
Joint fits of pn + MOS spectra. Errors at $1\sigma$. \\
$^{a}$ The blackbody normalization  corresponds to an emission radius of $177 d_4$ m during the flare and  $55 d_4$ m in quiescence.
\end{table*}

Considering the limited count statistics, we extracted the source spectra during the flare  using a maximum likelihood (ML) method, as described in \citet{2018arXiv180209454R}. 
The spectra obtained with the pn (130 net counts) and with the sum of the two MOS (121 net counts) were fitted simultaneously. 
An absorbed  power-law gave a good fit with photon index $\Gamma = 1.7\pm0.2$, absorption N$_H =(6\pm2)\times10^{21}$ cm$^{-2}$, and absorbed flux of $2.7\times10^{-12}$ erg cm$^{-2}$ s$^{-1}$  (0.5-10 keV).  
Acceptable fits were obtained also with other single component models (blackbody, thermal bremsstrahlung, thermal plasma emission,  multi blackbody disk emission), with the best fit parameters summarized in Table~\ref{tab-spectra}.

With the ML method it was possible to extract also the spectra of \src during the quiescent emission, i.e. from the whole observation excluding the time interval of the flare. This yielded exposure times of 6.5 ks in the pn and 8.8 ks   in the MOS  (100 and 85 source counts, respectively).
To avoid contamination from a nearby persistent source   (at only 18$''$ from \src , see Figure~\ref{fig-ima}), we included it in the ML model and extracted simultaneously the spectra of the two sources.  In the spectral fits of  \src , we fixed the absorption at the values derived from the spectrum of the flare with the corresponding model. We found a quiescent source flux  slightly smaller than $10^{-13}$ erg cm$^{-2}$ s$^{-1}$ and, although the uncertainties are large, some evidence that the spectrum during the quiescent period is softer than that of the flaring emission (see Table~\ref{tab-spectra}).  

 \subsection{\cha}
 
  \ngc   was observed with the \cha  ACIS instrument \citep{2003SPIE.4851...28G}  for   5.1 ks on 2008 November 1. The target was imaged on the backside-illuminated Chip S3 of the ACIS-S array. 
A source detection,  
in the 0.3-8 keV energy range, 
revealed several sources in the cluster region, including one  positionally  coincident with  \src  and with a   count rate consistent with that observed for its quiescent level with \xmm .
   
\begin{figure}[ht!]
	\centering
\includegraphics[width=7.3cm]{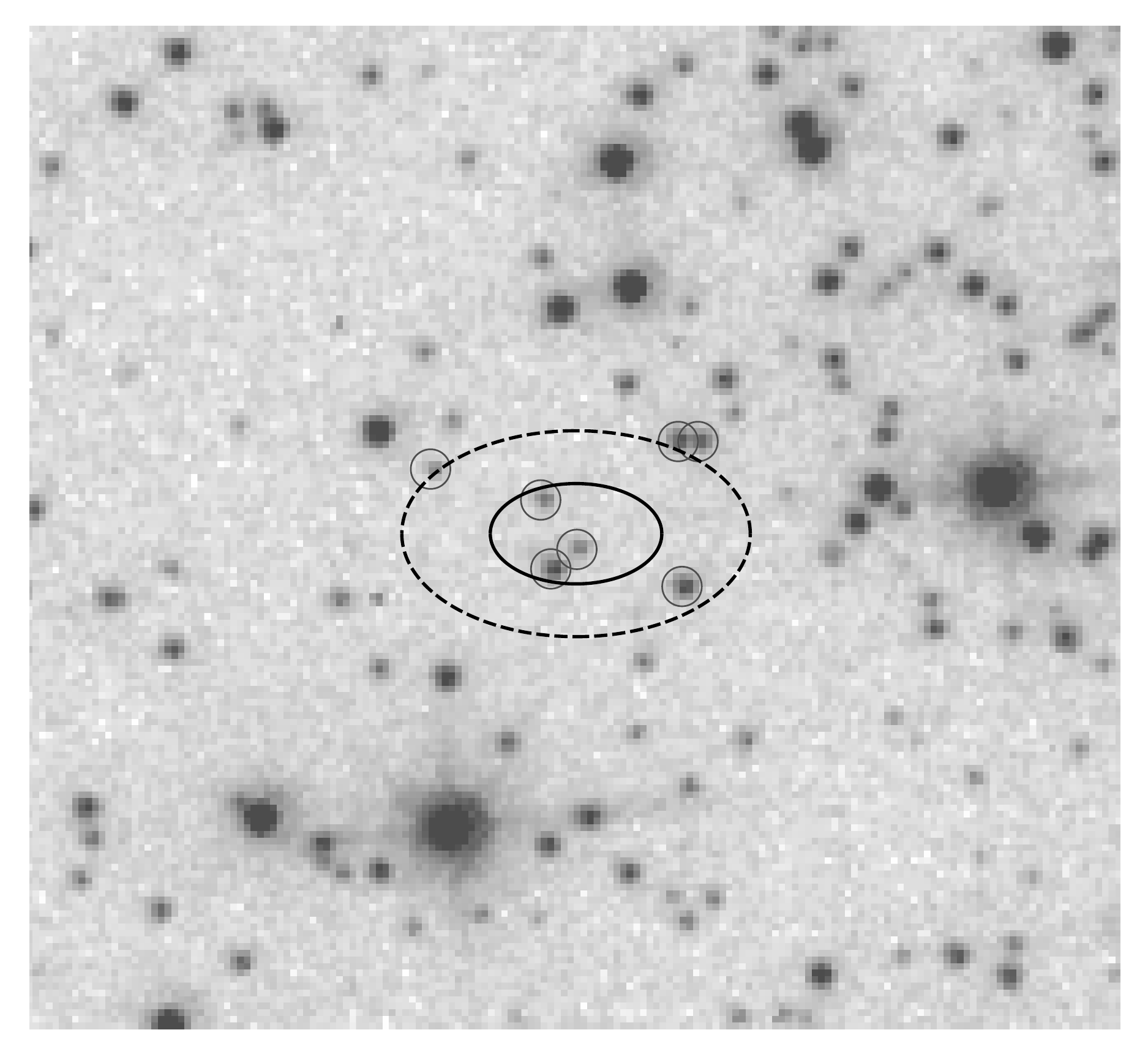}
	\caption{ Optical image of a  $\sim$8$''\times8''$ region centered at the position of \src obtained with 
the Planetary Camera in the F555W filter. North is to the top, East to the left. The error region of the transient source \src derived from the  \cha data is overplotted; the inner and outer ellipses  mark the 68\% and 99\% uncertainty region, respectively (see text for more detail). The seven HST sources closest to the \cha coordinates are marked; their position in the colour-magnitude diagram is shown in Figure~\ref{fig-cm}. 
\label{fig-hst}}
 \end{figure}

We improved the \cha astrometry by cross-correlating X-ray sources with the 2MASS catalog (also used as a reference for HST astrometry, see next section). 
We used 25 \cha sources located within 4 arcmin of the aimpoint, where the sharper point spread function yields a better localization accuracy. We found five matches within a correlation distance of 0.5$''$ -- no more matches were found up to distances larger than 1$''$. 
In view of the 2MASS source density of $\sim8.3\times10^{-3}$  sources per square arcsec, such five matches have a chance occurrence probability of $\sim8\times10^{-7}$ -- these are very likely the actual X-ray counterparts of the 2MASS sources. Adopting the five sources as a reference, we computed the best position of \src to be R.A. = 18$^h$ 06$^m$ 09.11$^s$, Dec. = --27$^{\circ}$ 45$'$ 54.8$''$. The $1\sigma$ error ellipse, shown by the solid line in Figure~\ref{fig-hst}, has semi-axis of 0.65$''$ and 0.38$''$ along R.A. and Dec., respectively. They include the uncertainty of the registration of both the HST and the \cha image to the 2MASS reference frame, as well as the statistical uncertainty on the localization of the X-ray source.  The \cha position is at a distance of 7$''$ from the center of the cluster (R.A. = 18$^h$ 06$^m$ 08.6$^s$, Dec. = --27$^{\circ}$ 45$'$ 55$''$, \citealt{1987ApJ...317L..13D}).

 \subsection{HST}
  
\begin{figure}[ht!]
\includegraphics[angle=0,width=7cm]{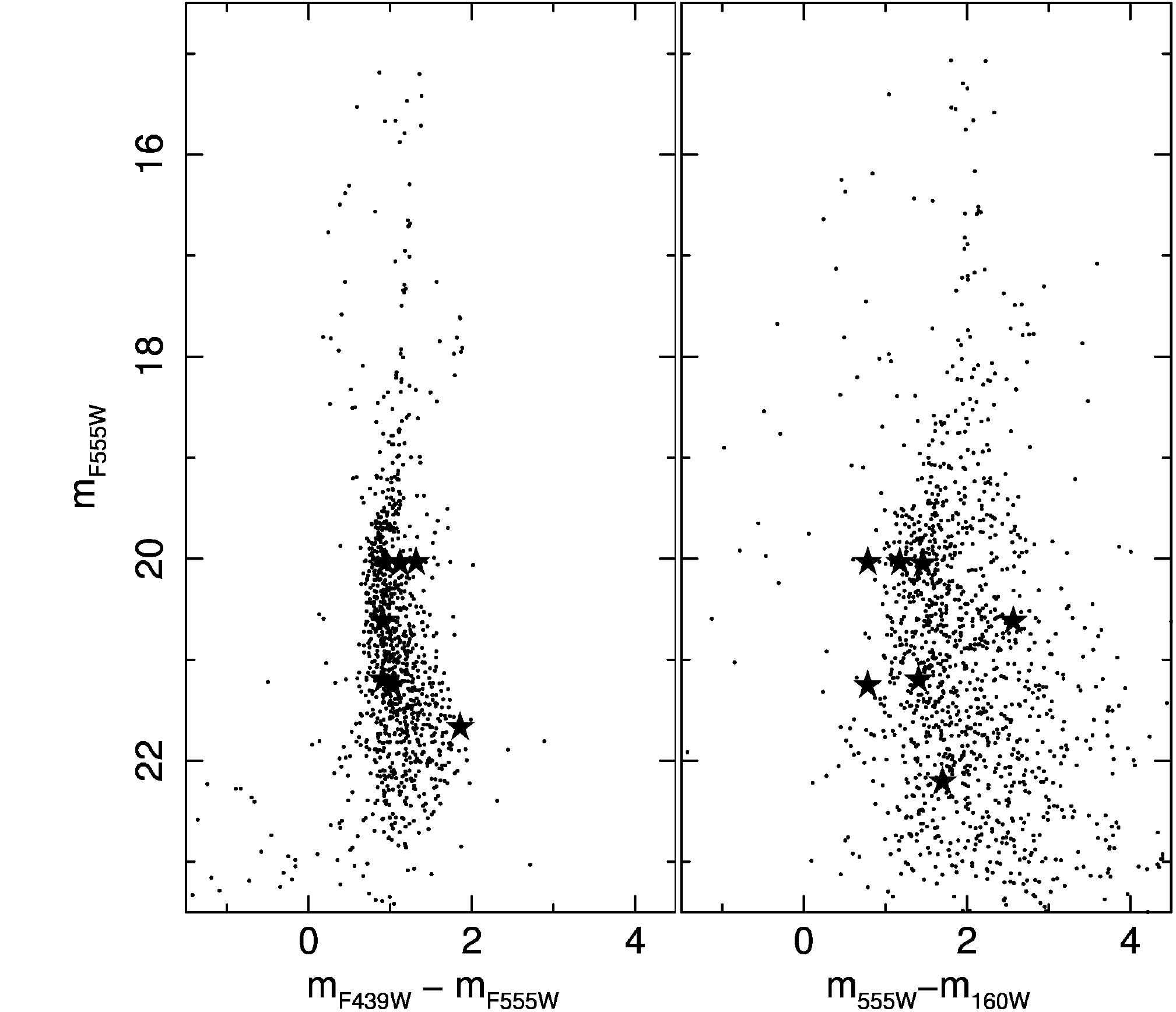}
	\caption{ Colour-magnitude diagrams for the cluster \ngc .
	 Magnitudes are in the AB system and are not corrected for reddening. Seven sources positionally consistent with the \cha error ellipse of \src  are indicated by the stars.
\label{fig-cm}}
 \end{figure}

The field of \ngc was visited by the {\em Hubble} Space Telescope on 2000, May 18 within a snapshot survey of Galactic globular clusters \citep[Prop. Id. 8118,][]{2002A&A...391..945P}. Short observations were performed with the Wide Field and Planetary Camera 2 (WFPC2) instrument in the F439W band ($\lambda=4311$ \AA, $\Delta\lambda=473$ \AA) and F555W band ($\lambda=5439$ \AA, $\Delta\lambda=1228$ \AA), with exposure times of 360 s and 45 s, respectively.  More recently (2016, July 19), the field was observed in the near infrared with the Wide Field Camera 3 (WFC3) instrument in the F110W band ($\lambda=11534$ \AA, $\Delta\lambda=1428$ \AA) and F160W band ($\lambda=15369$ \AA, $\Delta\lambda=826$ \AA), with exposure times of 1240 s and 1270 s, respectively. We retrieved calibrated, geometrically-corrected images from the {\em Hubble} Legacy Archive (HLA\footnote{https://hla.stsci.edu/hlaview.html}). HLA WFPC2 images have an accurate astrometry, based on cross-correlation of sources with astrometric catalogues -- the images of our field have a r.m.s. accuracy of $0.17''$ per coordinate based on 176 reference sources in 2MASS.  Such an accurate astrometry is not provided  for WFC3 images in the HLA. We  computed a more refined astrometric solution by cross-correlating WFC3 sources with the 2MASS catalog, with a r.m.s. accuracy of $\sim0.16''$ per coordinate, based on more than 100 reference sources. We ran a source detection 
 using the SExtractor software \citep{1996A&AS..117..393B}
and   cross-correlated the results of the two bands to generate a common source list. We performed  aperture photometry using $0.3''$ radii and applying standard aperture correction \citep{WFPC2HB}. We converted count rates to magnitudes in the AB system by using the photometric information provided by the 
HLA pipeline.   Colour-magnitude diagrams based on the images in the F439W, F555W and F160W filters, not corrected for reddening, are shown in Figure~\ref{fig-cm}, where sources positionally consistent with the \cha error ellipse are highlighted.
  
\section{Discussion}
 
The fluxes of the variable source \src derived in the previous section, corrected for the interstellar absorption, imply a quiescent X-ray  (0.5--10 keV) luminosity of  $L_{\rm Q}=(0.6-2.3)\times10^{32} d_4^2$ erg s$^{-1}$ and an average luminosity during the flare of   $L_{\rm F}=(4-8)\times$10$^{33} d_4^2$ erg s$^{-1}$,  where the quoted intervals account for the statistical errors and for the uncertainties on the spectral shape.

If \src is located in \ngc , the flare luminosity is much smaller than that typically seen during the outbursts of  LMXB transients.
Furthermore, such outbursts have much longer durations, lasting at least a few days and, more often, several weeks \citep[see, e.g.,][]{1998A&ARv...8..279C}. 
Besides the ``classical'' LMXB transients, there is a class of so called  ``very-faint X-ray transients''  (VFXTs) that show  fainter outbursts, reaching  peak   luminosities of only 10$^{34-36}$ erg s$^{-1}$   (e.g. \citet{2009A&A...495..547D}). The occurrence of type I bursts, shows that at least some VFXTs are accreting neutron stars in LMXBs, but also other interpretations have been proposed.  Considering the uncertainties on the distance of \ngc, the peak luminosity reached during the flare of  \src is marginally consistent with the range observed in VFXTs,  which however show outbursts of much longer duration.

The short duration of the  flare observed in \src is  more reminiscent of a  type I burst. However, type I bursts have a much higher luminosity ($\sim10^{37-38.5}$ erg s$^{-1}$) and their light curves are characterized by a sharp rise ($\lsim$10 s) followed by a longer decay during which the spectrum softens \citep[see, e.g.,][]{2008ApJS..179..360G}. These properties are at variance with those of \src and  point against this interpretation. 

Our results on \src bear some resemblance with those obtained during an  ASCA satellite observation of the globular cluster M28  by  \citet{1997ApJ...490L.161G}. These authors observed a short flare with  peak luminosity of $4\times10^{36}$ erg s$^{-1}$ and  interpreted it as a peculiar  type I burst. They explained the unusually low luminosity as due to emission from a very small region (0.1--1\%) of the neutron star surface, as it would be expected in the presence of a magnetic field higher than that of normal X-ray bursters.  The event seen in M28 differs from that  we observed  in \src  because it showed a very fast rise time ($\lsim$70 ms) and an exponential decay with characteristic time $\tau$=7.5 s.  On the contrary, the light curve of the \src flare has a fairly symmetric profile, well fitted by a Gaussian  with $\sigma=54\pm11$ s (see inset of Figure~\ref{fig-lc}).

Globular clusters contain also a large population of millisecond pulsars and cataclysmic variables. The quiescent luminosity of \src is consistent with the values observed in these classes of objects \citep{2009ASSL..357..165G,2006csxs.book..421K}, which, however, do not show flares or outbursts with the characteristics seen in our data. Furthermore, none of the possible optical  counterparts of \src highlighted in Figure~\ref{fig-cm} has a particularly blue color.

In summary,  the   properties of the new transient \src are at variance with those observed up to now in   all the classes of neutron stars and white dwarf sources present in globular clusters. In the following, we consider  the alternative explanation in terms of a stellar X-ray flare.

The brightest stellar X-ray flares, reaching peak   luminosities up to  10$^{33-34}$ erg s$^{-1}$,   are emitted by RS CVn stars. RS CVn are binary systems, with orbital periods shorter than 10 days,  formed by a giant or subgiant  G or K type star  plus a late type main sequence or subgiant star. 
A fast rotation rate is believed to be at the origin of their large   coronal X-ray luminosity, typically in the range   10$^{29-32}$ erg s$^{-1}$ \citep{1981ApJ...245..671W,1993ApJS...86..599D}. 
BY Dra type stars are another related class of chromospherically active binaries, in which the two components are late type dwarfs. They  have, on average,  slightly lower X-ray luminosity than the RS CVn's. 
X-ray sources associated with  RS CVn and BY Dra systems have been found in globular clusters (e.g.,   in  NGC 6752, \citealt{2002ApJ...569..405P}; in $\omega$ Cen,  \citealt{2003A&A...400..521G};   in 47 Tuc, \citealt{2003ApJ...596.1177E}).

 \begin{figure}[ht]
	\centering
\includegraphics[angle=0,width=7.5cm]{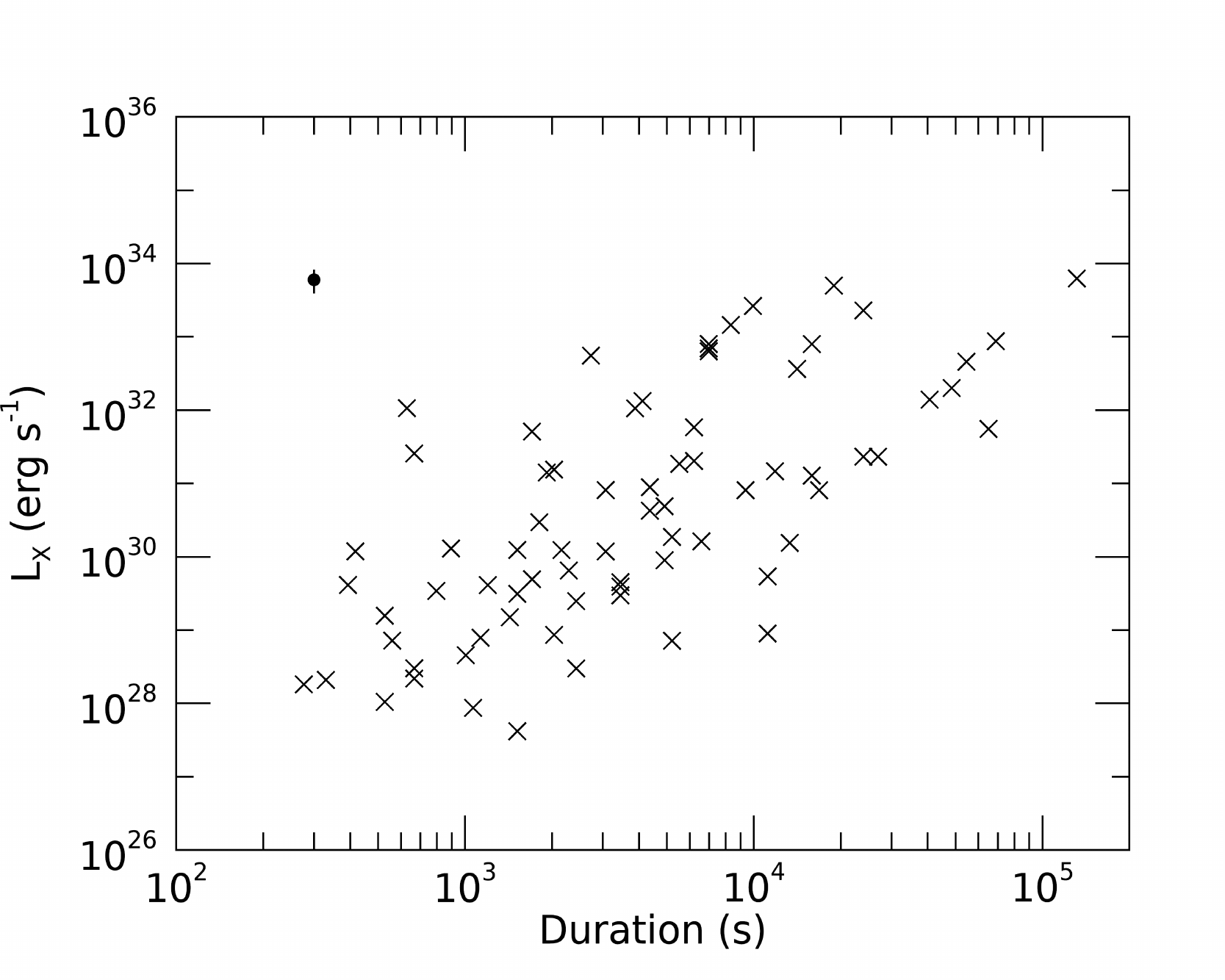}
			\caption{Comparison of duration and luminosity flare from \src (black dot) with those of a sample of flares from cromospherically active stars \citep[][ , and references therein]{2016PASJ...68...90T}. 
\label{fig-stars}}
 \end{figure}
 
The  values of  $L_{\rm Q}$ of \src , as well as the apparent magnitudes and colors of its possible optical counterparts,  are consistent with an active binary in \ngc .  
For example, a K-type dwarf with absolute magnitude  $M_V$ = 6 and intrinsic color $(V-R)_0\sim0.7$, at a distance of 4 kpc and with a reddening corresponding to that of \ngc ($E_{B-V}=0.7$) would have an apparent  magnitude $m_V\sim21$.   
As another comparison, we refer to the HST photometry of the faint X-ray sources associated to active binaries in 47 Tuc   \citep{2003ApJ...596.1177E}, which has a distance (4.5 kpc) similar to that of \ngc . They have optical properties spanning a broad region of the CMD which is consistent with that of the  candidate counterparts of \src .

The interpretation of the observed variable emission as a stellar flare leads us to adopt the spectral results obtained with the thermal bremsstrahlung fit (see Table~\ref{tab-spectra}). The flare average luminosity corresponds to an emission measure of $4\times10^{56} d_4^2$ cm$^{-3}$. This is rather high, but still compatible with the values observed in strong flares from RS CVn stars \citep{2016PASJ...68...90T}.
On the other hand the flare from \src  is particularly luminous in relation to its short duration, as can be seen in Figure~\ref{fig-stars}, where these properties are plotted for a large sample of stellar flares recently compiled by  \citet{2016PASJ...68...90T}. 
All the flares lasting less than 600 s have luminosities in the range 10$^{28}$--10$^{30}$ erg s$^{-1}$.
There is a correlation between the X-ray luminosity and duration, $\tau$, of the flares. When also solar flares and microflares are included, such a correlation holds over 12 orders of magnitude in luminosity, and gives $\tau \propto$ L$_X^{0.2}$ (see Fig. 5 of \citealt{2016PASJ...68...90T}).  This would predict a  luminosity of  a few 10$^{25}$ erg s$^{-1}$  for a flare with $\tau$=300 s,  much fainter than that observed from \src .

Of course, a much better agreement with the duration-luminosity relation for stellar flares would be obtained if \src  were a foreground source, not related to \ngc . However, this possibility is  disfavoured by the faintness  and colours of the optical candidate counterparts. For example, a flare with typical X-ray luminosity  of $10^{29-30}$ erg s$^{-1}$  \citep{2015A&A...581A..28P}  would require a star  closer than $\sim$50 pc.  
A typical dMe star (M$_V$$\sim$12) would appear brighter than $m_V\sim$16. A more extreme case of a  M8V star (M$_V\sim$16) at this close distance would give an apparent magnitude $m_V\sim$20, but it would have a color significantly redder than those of the possible counterparts plotted in Fig.~\ref{fig-cm}.  Furthermore, the absorbing column density derived from the X-ray spectra would be difficult to reconcile with that expected for a source at such a small distance. We therefore believe that the explanation in terms of a foreground flaring M star is very unlikely.

 \section{Conclusions}

The  variable source \src , most likely located in the Galactic globular cluster \ngc ,  was discovered thanks to the  emission of a brief X-ray flare during a systematic search for variability in archival \xmm data. 
The properties of the flare, in particular its short duration (300 s), symmetric time profile, and low luminosity ($(4-8)\times$10$^{33} d_4^2$ erg s$^{-1}$), are significantly different from those typically shown by outbursts and/or type I bursts from globular cluster X-ray sources containing compact objects. An interpretation in terms of a stellar flare seems more plausible, but also in this case the source is quite unusual because flares of such a high X-ray luminosity are expected to have a much longer duration.

It is possible that the unusual properties of this event reflect the fact that the EXTraS project is sampling for the first time with adequate sensitivity  a  region of the  parameter space relatively unexplored. Indeed, short flares from relatively weak and poorly sampled sources are difficult to discover and it is interesting to note that only  very  few   events as short as that of \src were found in the extended analysis carried out in the EXTraS project 
(encompassing more than 400,000 source detections listed in the  3XMM catalog, and also searching for new, undetected transients).   It is remarkable that one of such short flares  was found in a source positionally coincident with the central part of a globular cluster.
 
Further investigation of the candidate optical counterparts can shed light on the nature of \src , possibly confirming the dentification with a chromospherically active binary. However, also considering the large number of compact objects present in globular clusters other more exotic explanations (e.g.  some form of flaring magnetar-like emission from a neutron star, or a peculiar outburst from a compact object) cannot be excluded.

\begin{acknowledgements}

\src  was selected as a potentially interesting source by L. Apollonio, B. Bottazzi-Baldi, M. Giobbio, R.~F. Patrolea, E. Pecchini and C.~A. Torrente (Liceo Scientifico G.~B. Grassi, Saronno) during their stage at INAF-IASF Milano in 2017, September, within the {\em Alternanza Scuola-Lavoro} initiative of the Italian Ministry of Education, University and Research.

We acknowledge financial support from the Italian Space Agency (ASI) through the ASI-INAF agreements 2015-023-R.0 and 2017-14-H.0.

This research has made use of data produced by the EXTraS project, funded by the European Union's Seventh Framework Programme under grant agreement n. 607452. The EXTraS project acknowledges the usage of computing facilities at INAF -  Astronomical Observatory of Catania. The EXTraS project acknowledges the CINECA award under the ISCRA initiative, for the availability of high performance computing resources and support.

This work is partly based on observations made with the NASA/ESA Hubble Space Telescope, and obtained from the Hubble Legacy Archive, which is a collaboration between the Space Telescope Science Institute (STScI/NASA), the Space Telescope European Coordinating Facility (ST-ECF/ESA) and the Canadian Astronomy Data Centre (CADC/NRC/CSA). We also used observations made by the Chandra X-ray Observatory, and obtained from the Chandra Data Archive.

\end{acknowledgements}

\end{document}